\documentclass[%
 reprint,
 superscriptaddress,
 amsmath,amssymb,
 aps,
 prr,
longbibliography,
]{revtex4-2}

\usepackage{graphicx}
\usepackage{dcolumn}
\usepackage{bm}
\usepackage{physics}
\usepackage{hyperref}
\usepackage{xparse}
\usepackage{color}
\usepackage{appendix}

\NewDocumentCommand{\rnl}{O{r} O{n} O{l}}{X_{#2 #3} (#1)}

\newcommand{\MIS}{\bar{Z}}
\newcommand{\RWS}{R_\textrm{VS}}
\newcommand{\Ne}{N_\textrm{e}}

\renewcommand{\vec}{\mathbf}
\newcommand{\change}[1]{\textcolor{black}{#1}}

\begin{document}

\preprint{APS/123-QED}

\title{Improved calculations of mean ionization states with an average-atom model}

\author{Timothy J. Callow}
\email{t.callow@hzdr.de}
\affiliation{Center for Advanced Systems Understanding (CASUS), D-02826 Görlitz, Germany}
\affiliation{Helmholtz-Zentrum Dresden-Rossendorf, D-01328 Dresden, Germany}
\author{Eli Kraisler}%
 \email{eli.kraisler@mail.huji.ac.il}
\affiliation{Fritz Haber Center for Molecular Dynamics and Institute of Chemistry, The Hebrew University of Jerusalem, 9091401 Jerusalem, Israel}


\author{Attila Cangi}
 \email{a.cangi@hzdr.de}
\affiliation{Center for Advanced Systems Understanding (CASUS), D-02826 Görlitz, Germany}
\affiliation{Helmholtz-Zentrum Dresden-Rossendorf, D-01328 Dresden, Germany}



\date{\today}

\begin{abstract}


The mean ionization state (MIS) is a critical property in dense plasma and warm dense matter research, for example as an input to hydrodynamics simulations and Monte--Carlo simulations. Unfortunately, however, the best way to compute the MIS remains an open question. Average-atom (AA) models are widely-used in this context due to their computational efficiency, but as we show here, the canonical approach for calculating the MIS in AA models is typically insufficient. We therefore explore three alternative approaches to compute the MIS. Firstly, we modify the canonical approach to change the way electrons are partitioned into bound and free states; secondly, we develop a novel approach using the electron localization function; finally, we extend a method which uses the Kubo--Greenwood conductivity to our average-atom model. Through comparisons with higher-fidelity simulations and experimental data, we find that any of the three new methods usually out-performs the canonical approach, with the electron localization function and Kubo--Greenwood methods showing particular promise.

\end{abstract}

\maketitle





\section{Introduction}

Warm dense matter (WDM) is a phase of matter characterized by temperatures on the order of $1-100$~eV and densities of $10^{-2}-10^4\ \textrm{g cm}^{-3}$  \cite{DOE09,bonitz_review_2020}. Under these conditions, conventional divisions between solid-state and plasma physics are bridged and a variety of interesting phenomena emerge, including for example non-equilibrium effects \cite{non_eq_1}, phase transitions \cite{phase_transition_1,phase_transition_2}, and partially ionized matter. WDM is observed in various astrophysical domains, such as exoplanets \cite{exo_1} and brown and white dwarfs \cite{brown_dwarves,white_dwarves}; furthermore, during inertial confinement fusion (ICF), materials are exposed to WDM conditions \cite{ICF,Glenzer_adiabats}.

The mean ionization state (MIS), or equivalently the free electron density, is of particular importance in WDM. It is directly related to physical properties such as electrical conductivity, opacity, collision rates and acoustic velocities \cite{Redmer_Kubo_Greenwood_20,Glenzer_XRTS_1999}. Furthermore, the MIS is an input parameter for various simulations including hydrodynamics \cite{Murillo_hydro_2016} and Monte--Carlo simulations \cite{Vorberger_ionion_2013}, finite-temperature pseudo-potentials for density-functional theory calculations \cite{Perrot_pseduo_1995,Dharma_wardana_pseudo_2006}, and in computing adiabats used in ICF modelling \cite{Glenzer_adiabats}.
Additionally, accurate predictions of the MIS are crucial for validating and fitting models to experimental data \cite{Lithium_IPD_expt_2008,Kraus_Carbon_IPD_2018}.

In the WDM regime, it is often difficult to distinguish between `bound' and `free' electrons, meaning the MIS is hard to define. The ramifications of this ambiguity extend beyond direct computation of the MIS: they are relevant to recent debates regarding the ionization potential depression (IPD) effect \cite{Hu_IPD_2017,Iglesias_plea_2014,Sterne_comment_2018,Hu_reply_2018}, and further raise questions regarding the application of the Chihara decomposition \cite{Chihara_1987,Chihara_1999, Baczewski_nochihara_2016}. These difficulties are further compounded by the variety of methods used in the modelling of WDM, running all the way from analytical models such as Stewart--Pyatt \cite{SP66} and Ecker--Kroll \cite{EK63} to \emph{ab initio} density-functional theory (DFT) \cite{HK64,KS65,M65,D03,HRD08} and path-integral Monte--Carlo \cite{DM12,DGB18} simulations. It is therefore of great interest to develop an approach for calculating the MIS that is consistent between different models and experimental results.

Average-atom (AA) models are a popular and successful tool in modelling the WDM regime, since they incorporate in a natural way quantum effects (typically using DFT) at a manageable computational cost \cite{FMT49,R72,Li_1979}. There is a wide range of AA models \cite{callow_first_principles}, but they share in common the concept of an atom immersed in a plasma. Typically, the MIS is defined as the number of electronic states with energy above a certain threshold,
\begin{align} \label{eq:MIS_thresh}
    \MIS &= \int_{\epsilon_0}^\infty \dd{\epsilon} g(\epsilon)  f_\textrm{FD} (\epsilon)\,,
\end{align}
where $g(\epsilon)$ denotes the density-of-states, $f_\textrm{FD} (\epsilon)$ the Fermi--Dirac (FD) distribution, and $\epsilon_0$ the chosen energy threshold. In AA models, the threshold energy is typically chosen to be the value of the mean-field potential at the boundary of the Voronoi cell $\RWS$ (the atomic radius), $\epsilon_0 = v_\textrm{s}(\RWS)$. Other choices for $\epsilon_0$, for example equating it to the chemical potential, could also be considered.

As seen in a previous work \cite{callow_first_principles}, the definition \eqref{eq:MIS_thresh} is somewhat limited, showing large discrepancies for different choices of boundary condition and sharp discontinuities when $\MIS$ is plotted as a function of temperature or density. Furthermore, bound and free states in AA models are typically treated differently (although not always, for example Refs. \cite{Son_IPD_AA_2014,Massacrier_bands_2021}): the definition of $\MIS$ is thus both an output of and input to the model, which means any errors may self-multiply.

DFT-based molecular dynamics (DFT-MD) simulations can also be used to compute the MIS using definition \eqref{eq:MIS_thresh}, with the threshold energy typically assumed to start at the conduction band lower edge, $\epsilon_0=\epsilon_\textrm{c}$. In DFT-MD simulations, all (non-core) orbitals are treated on the same footing, which is an advantage (among others) relative to AA models. However, there are still (at least) two limitations using this definition, which are common to both DFT-MD and AA models. The first is the ambiguity about how to define the threshold energy. The second is the assumption that states can be categorized as completely bound or completely free based on their energy alone: DFT-MD results with this method have shown counter-intuitive behaviour \cite{Priesing_Hellium} and divergence from experimental measurements \cite{Lithium_IPD_expt_2008}.  

Consequently, novel ways of computing the MIS have recently gained traction. For example, Bethkenhagen~\emph{et al.} proposed using the Kubo--Greenwood (KG) conductivity formula to measure the MIS \cite{Redmer_Kubo_Greenwood_20}. This approach was applied to Carbon under high temperatures and gigabar pressures (and later to the metallization of helium \cite{Priesing_Hellium}), and the resulting MIS values showed disagreement with various other methods. Interestingly, excellent agreement was seen between pressures computed with an AA model and DFT-MD under these conditions \cite{Carbon_ionization_Faussurier_2021}. However, the MIS computed with the same AA model --- using yet another definition for $\MIS$ --- had a systematic error relative to the DFT-MD KG result, which suggests that a more pertinent definition of the $\MIS$ in an AA model might give better agreement.

In this paper, we explore three methods for computing the MIS in an AA model, and compare results with DFT-MD simulations \cite{Redmer_Kubo_Greenwood_20} and experimental data \cite{Glenzer_Beryllium_expt_2003,Vinko2015,Aluminium_IPD_expt_2012}. Firstly, we apply the canonical definition \eqref{eq:MIS_thresh}, which (as expected) gives inconsistent results, particularly for high densities. Secondly, we modify the canonical approach such that the orbitals are no longer categorized as bound or free based on their energy. Instead, they are partitioned depending on their shell ($1s$, $2s$, etc), \change{an approach that was used for the (non-average-atom) XCRYSTAL model in Ref.~\onlinecite{xcrystal}}. Thirdly, we introduce a novel approach which uses the electron localization function (ELF) to determine the MIS. The ELF is well-known in quantum chemistry and materials science \cite{Savin_ELF_1996, Fuentealba_understanding_ELF_2007}, but has not until now been applied to study ionization in WDM. We shall see that this method yields more consistent and accurate results compared to the canonical approach. Finally, we adapt the KG method of Ref.~\onlinecite{Redmer_Kubo_Greenwood_20} to our AA model. This approach shows excellent agreement both with DFT-MD simulations and the experimental results, but is so far limited to only one boundary condition in the AA model. Nevertheless, the ELF and KG results demonstrate that computationally efficient AA models can accurately and reliably predict the MIS across a wide range of conditions.


\section{Theory}

\subsection{Average-atom model}

The AA model we use is a generalization of the model derived in Ref.~\onlinecite{callow_first_principles}. 
We explain here the main features of this model and the differences from the one presented in Ref.~\onlinecite{callow_first_principles}; however, we direct readers to that paper for a detailed derivation and discussion  of this AA model.
In our AA model, we solve the Kohn--Sham DFT (KS-DFT) equations for a single atom consisting of a nucleus with charge $Z$ and a fixed number of electrons $\Ne$ (with $\Ne=Z$ for all the systems we consider). Explicit interactions between this atom and its neighbours are ignored, and instead these interactions are implicitly accounted for via the boundary conditions imposed on the orbitals at the sphere's edge (Voronoi sphere radius, $\RWS$).

The spherically symmetric KS equations to be solved are given by \footnote{\change{We note here one difference from the AA model presented in Ref.~\cite{callow_first_principles}: in this paper, we solve the spin-unpolarized KS
equations, i.e. the spatial spin-up and spin-down orbitals
are assumed to be identical,
\(X_{nl}^\uparrow(r)=X_{nl}^\downarrow(r)\). }}
\begin{multline}
\left[\frac{\textrm{d}^2}{\textrm{d}r} + \frac{2}{r}\frac{\textrm{d}}{\textrm{d}r} - \frac{l(l+1)}{r^2} \right] X_{nl}(r)  \\
+ 2 \left[\epsilon_{nl} - v_\textrm{s}[n](r) \right] X_{nl}(r) = 0, \label{eq:KSeqns}
\end{multline}
where \(v_\textrm{s}[n](r)\) is the KS potential, given by
\begin{equation}
 v_{\textrm{s}}[n](r) = -\frac{Z}{r} + 4\pi \int_0^{R_\textrm{WS}} \textrm{d}{x} \frac{n(x)x^2}{r^>(x)} + \frac{\delta F_\textrm{xc} [n]}{\delta n(r)}\,,
\end{equation}
with \(r^>(x)=\max(r,x)\). The three terms in the potential are
respectively the electron-nuclear attraction, the classical Hartree
repulsion, and the exchange-correlation (xc) potential, which is equal
to the functional derivative of the xc free energy. As ever, due to the
dependence of the KS potential on the density \(n(r)\), the KS equations
must be solved iteratively until self-consistency is reached.

The density \(n(r)\) is constructed from the orbitals as
\begin{equation}
n(r) = 2\sum_{nl}(2l+1) f_{nl}(\epsilon_{nl},\mu,T) |X_{nl}(r)|^2\,. \label{eq:KSdens}
\end{equation}
where \(f_{nl}(\epsilon_{nl},\mu,T)\) is the Fermi--Dirac (FD)
distribution, given by
\begin{equation}
f_{nl}(\epsilon_{nl},\mu,T) = \frac{1}{1+e^{(\epsilon_{nl}-\mu)/T}}\,.
\end{equation}we have not made any changes to avoid too much speculation.

The chemical potential \(\mu\) is determined by fixing the electron
number \(N_\textrm{e}=4\pi\int_0^{\RWS} \textrm{d}r r^2 n(r)\) to be equal to a
pre-determined value (in this paper, \(N_\textrm{e}=Z\) in all cases).

\change{We impose} boundary conditions on the KS
orbitals \(X_{nl}(r)\) which are intended to implicitly account for
inter-atomic interactions. In our earlier paper
\cite{callow_first_principles}, we argued that a physically intuitive
condition was to impose smoothness of the density at the edge of the \change{Voronoi sphere (VS)},
\begin{equation}
\frac{\textrm{d}n(r)}{\textrm{d}r}\Bigg|_{r=\RWS} =0\,.
\end{equation}
Mathematically there is no unique way to enforce the above condition,
but two simple choices are
\begin{align}
0&=X_{nl}(\RWS)\,, \\
0&=\frac{\textrm{d}X_{nl}(r)}{\textrm{d}r}\Bigg|_{r=\RWS}\,,
\end{align}
which we refer to respectively as ``Dirichlet'' and ``Neumann'' conditions. From a theoretical standpoint within the AA model, there is no way to unambiguously differentiate between these boundary conditions. 

We now note a key improvement we have made to our AA model compared to Ref.~\cite{callow_first_principles}. In that paper, we only solved the KS equations \change{\eqref{eq:KSeqns}} for the `bound' electrons, defined as those with energies below the threshold energy $\epsilon_0$. For the remaining `unbound' electrons, we used the ideal approximation, which amounts to assuming a constant density for the bound electrons, $n_\textrm{ub}(r)=\bar{n}$. In this work, we make \emph{no distinction} between `bound' and `unbound' orbitals during the SCF procedure: in other words, we solve the same equations \eqref{eq:KSeqns} for \emph{all} orbitals, regardless of their energy. As already mentioned, this removes the issue of the MIS being both an input to the model (via the ionization threshold $\epsilon_0$) and an output of it. Moreover, as we shall soon see, the expressions for the KG conductivity and the ELF are explicitly orbital-dependent; it therefore does not make sense to calculate these properties when part of the density is constructed in an orbital-free manner as we had done in the past.


Furthermore, to extend our comparisons beyond the model described above, we have implemented the AA model proposed by Massacrier \emph{et al.} \cite{Massacrier_bands_2021}. In this model, the KS equations are solved for \emph{both} the Dirichlet and Neumann boundary conditions, yielding energies $\epsilon_{nl}^\pm$ which define the upper (Dirichlet) and lower (Neumann) limits of a band-structure. Within these limits, every energy value is permitted and the wave-function corresponding to that energy is determined. The KS equations thus become
\begin{multline}
\left[\frac{\textrm{d}^2}{\textrm{d}r} + \frac{2}{r}\frac{\textrm{d}}{\textrm{d}r} - \frac{l(l+1)}{r^2} \right] X_{\epsilon nl}(r)\\ + 2 \left[\epsilon^{\tau}_{nl} - v_\textrm{s}[n](r) \right] X_{\epsilon nl}(r) = 0\,.
\end{multline}

The Fermi--Dirac occupations are multiplied by the Hubbard density-of-states (DOS) function $g_{nl}(\epsilon)$, defined as \cite{Hubbard_DOS}
\begin{gather}
g_{nl}(\epsilon) =\frac{8}{ \pi \Delta_{nl}^2} \sqrt{(\epsilon^+_{nl}-\epsilon)(\epsilon - \epsilon^-_{nl})}\,,\\
\Delta_{nl} = \epsilon^+_{nl}-\epsilon_{nl}^- \,,
\end{gather}
which means the density in this band-structure model is given by
\begin{equation}
    n(r) = 2\sum_{nl}(2l+1)
    \int_{\epsilon_{nl}^-}^{\epsilon_{nl}^+} \textrm{d}{\epsilon} g_{nl}(\epsilon) f_{nl}(\epsilon,\mu,\tau) |X_{\epsilon nl}(r)|^2\,.
\end{equation}

In practice, the energy bands are discretized, and the above integral becomes a summation over energies within each band which we now denote by index \(k\). Following some algebraic manipulation, the density can be written as
\begin{gather} \label{eq:dens_masac}
n(r) = 2\sum_{k}^{N_k} w_k \sum_{nl}(2l+1) f_{knl}(\epsilon_{knl},\mu,\tau) |X_{knl}(r)|^2\,,\\
w_k = \frac{8}{\pi(N_k-1)^2}\sqrt{k(N_k-1-k)}\,, \label{eq:weight_masac}
\end{gather}
where $N_k$ is the number of points used in the discretization of each energy band. The above expression closely resembles the expression for the density in plane-wave DFT codes, since it has a summation over \(k\)-points and
some weighting \(w_k\) (with $\sum_k w_k = 1$), very much like the \(\vec{k}\)-point mesh for
reciprocal space. It is also clear to see
that when the concept of bands in the AA model is not employed
(i.e.~when we use either the Dirichlet or Neumann conditions only), that
\(N_k=1,w_k=1\) and the above expression reduces to the ordinary expression for the density \eqref{eq:KSdens}. The above simplification (\ref{eq:dens_masac},\ref{eq:weight_masac}) was not shown in Ref.~\cite{Massacrier_bands_2021}, and we therefore provide a derivation in Appendix A.

\subsection{Counting method}

As discussed in the introduction, and we shall later see in the results, the canonical definition of the MIS in AA models \eqref{eq:MIS_thresh} is often erroneous, which is why we shall explore alternative approaches.
In the following three sub-sections, we discuss the application of new methods --- first, the counting method, which is a modification to the threshold approach, secondly, the electron localization function (ELF), and lastly, the Kubo--Greenwood conductivity --- to calculating the MIS. 

As seen in Eq.~\eqref{eq:MIS_thresh}, the canonical approach to computing the MIS essentially defines electrons as bound or free depending on whether their energy exceeds some threshold $\epsilon_0$, typically defined as the value of the KS potential at the edge of the atomic sphere. Intuitively, this does make sense, if one imagines electrons being bound so long as their energies are below the maximum value of the KS potential, and otherwise free. However, as we shall see in the Results section, this leads to unphysical discontinuities in the MIS when the energy of an orbital crosses the threshold value, and is very sensitive to the choice of boundary conditions.

Rather than making the bound-free partition dependent on some energy value, we instead propose to partition the electrons based on their shells. \change{This method was used to compute the MIS in Ref.~\onlinecite{xcrystal} for the XCRYSTAL model (specifically Eq.~(22) and the surrounding discussion), but has not been applied (as far as we know) to average-atom models. In fact, the argument they use for this approach --- ``our flat potential $V_0$ does not share the same physical interpretation as the flat potential used in Ref. [29], as delocalized
states can be found below $V_0$ in XCRYSTAL'' --- is applicable to average-atom models such as ours, in which there are no constraints on the KS potential.}

The method is perhaps best illustrated with an example. Consider Aluminium at its ambient density, $\rho_\textrm{m}=2.7\ \textrm{g cm}^{-3}$. It is well-known that, at room temperature, the $1s$, $2s$ and $2p$ orbitals are core states, and the remaining orbitals represent free electron density. This can also be seen by inspection of the density-of-states, using for example the average-atom band-structure model.

\begin{figure*}
    \centering
    \includegraphics{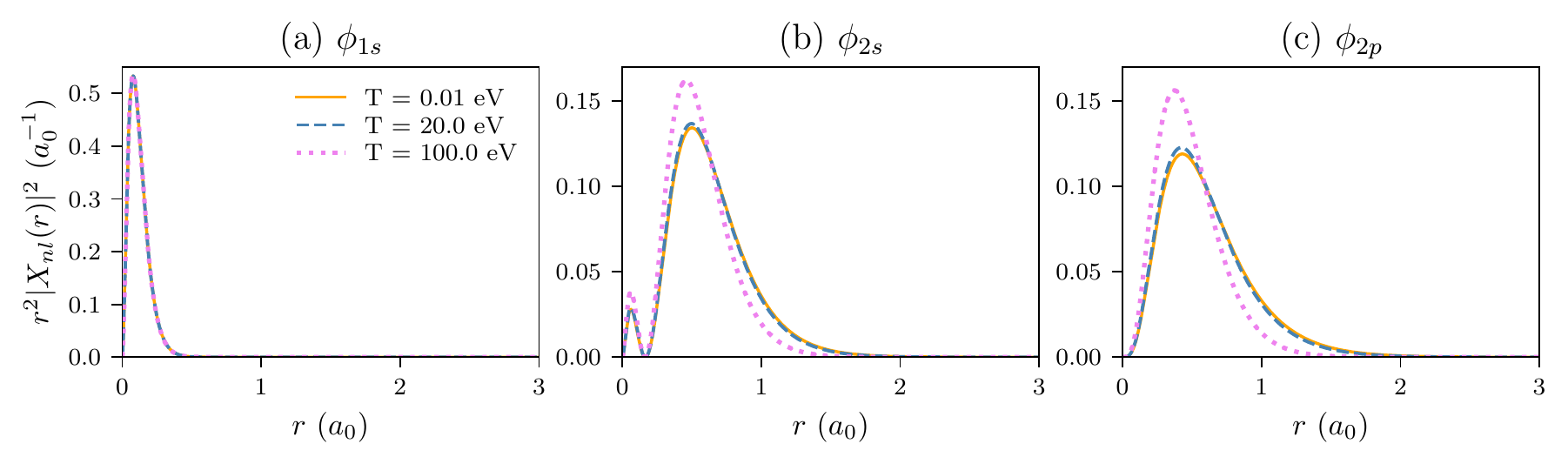}
    \caption{Radial KS orbitals (multiplied by $r^2$) for Aluminium at its ambient density (2.7 $\textrm{g cm}^{-3}$), for different temperatures. We see that the $1s$ state is unaffected by the temperature, and the $2s$ and $2p$ states are moderately affected, but not so much to change their bound-state character. The calculation was done with the Dirichlet boundary condition, but due to the core-nature of the orbitals, the boundary condition is of minimal impact.}
    \label{fig:Al_orbs}
\end{figure*}

As the temperature is increased, the character of these core states actually does not change much, as can be seen in Fig.~\ref{fig:Al_orbs}. Therefore we can essentially consider these states to represent bound electron density, regardless of the temperature. Of course, as the temperature increases, the occupation of these core states will decrease as higher-energy states are occupied, causing the MIS to increase. We shall henceforth refer to this approach as the `counting' method, and it can, in theory, be generalized to any material at a given density. The expression for the MIS in this counting method is
\begin{equation} \label{eq:Z_count}
    \MIS = \Ne - \sum_k w_k \sum_{n,l \in \textrm{b}} (2l+1) f_{nlk}\,,
\end{equation}
where $b$ denotes that subset of orbitals considered to be bound. Although we have included a $k$-dependence in the above sum, we have done so for generality; ideally, the bound states should be clearly identifiable as core states, in other words, their energies should not form a band and the $k$-index should be redundant.

Clearly, the approach described above works best if orbitals can be clearly identified as being of bound or free character, as is typical for metals at their ambient density (for example). However, when this is not the case --- in particular when a range of densities is spanned for a given material --- the above method is likely to break down. As material density changes, the orbital character also changes significantly, bands emerge and disappear, and so on. In such scenarios, one would expect this counting method to fail. In the Results section, we shall see that this expectation is borne out.

\subsection{Electron localization function}

In this subsection, we describe the method we have developed to compute the MIS with the electron localization function (ELF). The ELF has a long history in quantum chemistry \cite{Becke_ELF,Savin_ELF_1996,Fuentealba_understanding_ELF_2007} as a tool for understanding atomic structure and chemical bonding. It was originally conceptualized by Becke and Edgecombe \cite{Becke_ELF}, who supposed that the conditional probability density --- i.e., the probability of finding an electron at position $\vec{r}_1$ given another electron with the same spin at position $\vec{r}_2$ --- could be used as a basis to measure electron localization. It was later generalized by Savin \cite{Savin_ELF_1996} such that any spin-independent electron density could be considered. 

In KS-DFT, the expression for the (total density) electron localization function (ELF) is given by
\begin{align}\label{eq:elf}
    \textrm{ELF}(\vec{r}) &= \frac{1}{1 + [D(\vec{r})/D_0(\vec{r})]^2}\,,
\end{align}
where $D(\vec{r})$ and $D_0(\vec{r})$ are the electron pair density curvature (EPDC) functions for the system and for the uniform electron gas (UEG) respectively. These are given by

\begin{align}
    D(\vec{r}) &= \tau(\vec{r}) - \frac{1}{8}\frac{[\grad n(\vec{r})]^2}{n(\vec{r})},\\
    D_0(\vec{r}) &= \frac{3}{10} (3\pi^2)^{2/3} n^{5/3}(\vec{r})\,,
\end{align}
where $\tau(\vec{r})$ is the kinetic energy density. There are in fact multiple ways to define $\tau(\vec{r})$ \cite{Cohen_1979,Ayers_KED,Jiang_2020}, which of course all yield the total kinetic energy when integrated over all space. The definition most commonly adopted in the ELF is \cite{Fuentealba_understanding_ELF_2007}
\begin{equation}\label{eq:ked}
    \tau(\vec{r}) = \frac{1}{2}\sum_{i=1}^{\Ne} [\grad\phi_i(\vec{r})]^2\,.
\end{equation}

The motivation for the definition of the ELF \eqref{eq:elf} is to define electron localization in a quantitative manner, by using the EPDC of the UEG, a perfectly delocalized electron density, as a reference. The ELF is bounded in the range $0\leq \textrm{ELF}\leq 1$: a value of 1 indicates strongly localized electron density and a value of $1/2$ indicates equivalence with the (delocalized) UEG. 

\begin{figure}
    \centering
    \includegraphics{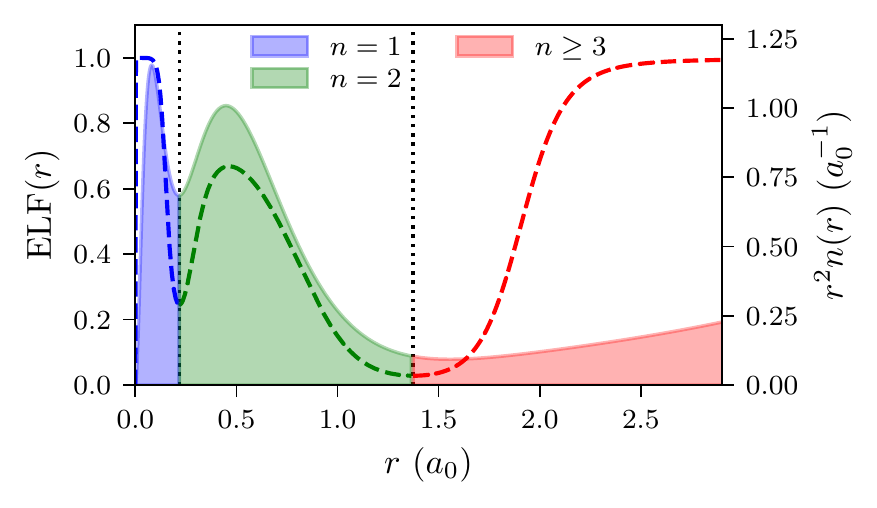}
    \caption{The ELF (dashed) and radial density distribution (shaded) for Aluminium at its ambient density $\rho_\textrm{m} = 2.7\ \textrm{g cm}^{-3}$ and low temperature ($\tau=0.01\ \textrm{eV}$). The figure shows how shells are defined via the minima of the ELF (positions emphasized with vertical dotted lines), with the electron number for that shell found by integrating the density in that region. \change{This figure corresponds to the orange line in Fig.~\ref{fig:ELF_big_comp}.(d), i.e. the Neumann boundary condition and definition \eqref{eq:ked} for the kinetic energy density.}}
    \label{fig:ELF_Al}
\end{figure}

One of the principal uses of the ELF is to calculate the number of electrons in particular shells. In the atomic picture, the spatial boundary of the shells is equated to minima in the ELF. Then, the density is integrated between minima to give the number of electrons in that shell. A visual example of this procedure is shown in Fig.~\ref{fig:ELF_Al}.

We propose to use the ELF as a measure of the MIS by computing the number of electrons per shell, and assuming that any electron density beyond a particular shell is free. This presents a similar issue to the counting method described in the prior sub-section; however, as we shall see, the ELF method is advantageous when a scan over densities is performed. Nevertheless, this does introduce some ambiguity and means this approach cannot be considered a ``black-box'' method. 


In the application of the ELF to our AA model at moderate-to-high temperatures, we have observed that, using the normal definition of the kinetic energy density \eqref{eq:ked}, the ELF's minima are often not identifiable. However, we have found that an approximate expression for the kinetic energy density $\tau(\vec{r})$, based on a second-order gradient expansion \cite{ELF_density_2002}, yields more clearly identifiable minima in the ELF than the normal orbital-dependent expression. This approximation for $\tau(\vec{r})$ is given by
\begin{multline} \label{eq:ked_approx}
    \tau(\vec{r}) = \frac{3}{10} (3\pi^2)^{2/3} n^{5/3}(\vec{r}) + \frac{1}{72}\frac{|\grad n(\vec{r})|^2}{n(\vec{r})} + \frac{1}{6}\grad^2 n(\vec{r})\,,
\end{multline}
which leads to the following expression for $D(\vec{r})$,
\begin{equation}
    D(\vec{r}) = D_0(\vec{r}) -\frac{1}{9} \frac{|\grad n(\vec{r})|^2}{n(\vec{r})} +  \frac{1}{6}\grad^2 n(\vec{r})\,.
\end{equation}
In spherical co-ordinates, this becomes
\begin{multline}
    D(r) = D_0(r) - \frac{1}{9}\frac{1}{n(r)}\left| \dv{n(r)}{r}\right|^2 \\
    +\frac{1}{6}\left(\frac{2}{r}\dv{n(r)}{r} + \dv[2]{n(r)}{r} \right)\,.
\end{multline}

\begin{figure*}
    \centering
    \includegraphics{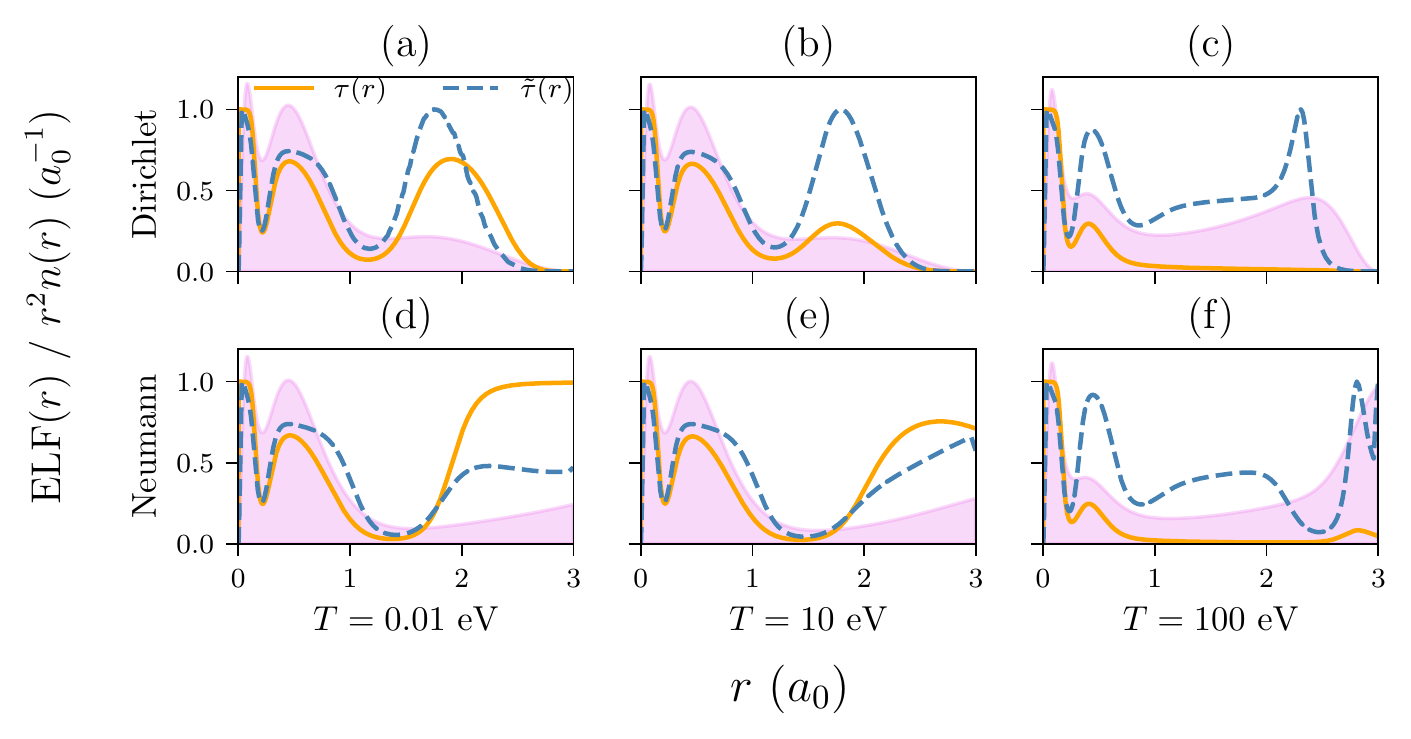}
    \caption{The ELF \change{(in orange and dotted blue lines)} and radial density distribution $r^2n(r)$ \change{(in shaded pink)} for Aluminium at its ambient density ($\rho_\textrm{m} = 2.7\ \textrm{g cm}^{-3}$), with different temperatures and boundary conditions. The ELF is computed in two ways, (i) using the definition \eqref{eq:ked} for the kinetic energy density $\tau(r)$ (solid orange line), and (ii) using the approximation \eqref{eq:ked_approx} $\tilde{\tau}(r)$ (dashed blue line). Although the minima for the $n=1$ and $n=2$ shells are both clearly visible at $T=0.01$ and $T=10$ eV using either method, at $T=100$ eV, the $n=2$ minimum can only be identified with the approximate definition \eqref{eq:ked_approx} of the kinetic energy density. \change{Panels (a)--(c) are calculated with the Dirichlet boundary condition and (d)--(f) with Neumann.}}
    \label{fig:ELF_big_comp}
\end{figure*}

In Fig. ~\ref{fig:ELF_big_comp}, we compare the ELF computed using the usual definition of the kinetic energy density \eqref{eq:ked} with the approximate form \eqref{eq:ked_approx}. We compare three temperatures: 0.01 eV, 10 eV and 100 eV, and consider both Dirichlet and Neumann boundary conditions. For 0.01 and 10 eV, we see that the shape of the ELF is in general different for the different forms of the kinetic energy density; however, the positions of the first two minima, which correspond to the boundaries of the $n=1$ and $n=2$ electron shells, are almost identical. On the other hand, at 100 eV, the $n=2$ minimum is no longer identifiable when the orbital-based definition \eqref{eq:ked} for the kinetic energy density is used; this is in contrast to when the approximate density-based definition \eqref{eq:ked_approx} is used, in which case the $n=2$ minimum is clearly visible. 

In general, we have observed the tendency for the orbital-based expression \eqref{eq:ked} to break down as temperature increases for a range of materials and densities. Consequently, we prefer to use the approximate definition \eqref{eq:ked_approx}, which does not display the \change{same} tendency, for all calculations of the MIS. As is observed in Fig.~\ref{fig:ELF_big_comp}, particularly in the right-hand panel of this figure, this can also produce additional and unexpected minima in the ELF. It is unclear whether these minima are really physically connected to electron shells, or are simply artifacts from the average-atom model and boundary conditions. Regardless, since we assume all electron density beyond a certain shell (\change{$n\geq 3$} in this example) is free, a correct physical interpretation of these additional minima is not strictly required in this approach.

\subsection{Kubo--Greenwood conductivity}

In this sub-section, we describe the application of the Kubo--Greenwood conductivity to compute the MIS within our AA model. The Kubo--Greenwood (KG) conductivity formula for a finite system is given by \cite{Johnson_KG,Trickey_Kubo_Greenwood_2017}
\begin{align}
        \sigma_{S_1,S_2}(\omega) = \frac{2\pi}{3V \omega} &\sum_{i\in S_1}\sum_{j\in S_2} (f_i - f_j) \nonumber |\mel{\phi_i}{\grad}{\phi_j}|^2 \\
        &\times\delta(\epsilon_j - \epsilon_i - \omega) \label{eq:kg},
\end{align}
where $\sigma(\omega)$ is the dynamical conductivity for two subsets $S_1$ and $S_2$ of the orbitals, $V$ is the volume of the system under consideration, $\phi_i$ are the KS orbitals and $\epsilon_i$ and $f_i$ are their energies and FD occupations. For the total conductivity, $S_1$ and $S_2$ represent the complete set of orbitals.

As described in Ref.~\cite{Redmer_Kubo_Greenwood_20}, Eq.~\eqref{eq:kg} can be used as a proxy for the mean ionization state in combination with the Thomas--Reiche--Kuhn (TRK) sum rule \cite{Thomas1925,Reiche1925,Kuhn1925}. This rule establishes a relationship between the KG conductivity and a certain number of electrons. For example, if we take $S_1$ and $S_2$ to both be the complete set of orbitals, then we should recover the total electron number,
\begin{equation} \label{eq:KG_Z_1}
    \Ne = \frac{2 V}{\pi} \int_0^\infty \dd{\omega} \sigma_{t,t} (\omega),
\end{equation}
\change{where $\sigma_{t,t}$ denotes the conductivity from the \textbf{t}otal, or complete, set of orbitals.} We note here that the complete set of orbitals means, in theory, an infinite set of KS orbitals (i.e. not just those with non-zero occupation numbers). In \change{practice}, a sufficient number of orbitals is chosen such that the resulting electron number is equal (within reasonable tolerance) to the expected electron number.  This provides a useful check of the implementation and convergence of the KG method.

To calculate the MIS, we use
\begin{equation} \label{eq:KG_Z_2}
    \MIS = \frac{2 V}{\pi} \int_0^\infty \dd{\omega} \sigma_{c,c} (\omega),
\end{equation}
where $\sigma_{c,c}$ means both orbital subsets are given by the conducting orbitals.

In the spherically symmetric AA model, the KG conductivity is given by
\begin{multline}
\sigma_{S_1,S_2}(\omega) = \frac{2\pi}{V\omega} \sum_{nl\in S_1} \sum_{n'l'\in S_2} \sum_{m\in \{S_1,S_2\}} (f_{nl} - f_{n'l'}) \\ |\nabla_{nn'll'm}^z|^2 \delta (\epsilon_{n'l'} - \epsilon_{nl} - \omega) \delta(l\pm 1 - l')\,,
\end{multline}
which leads to the following expression for $Z_{S_1,S_2}$,
\begin{multline}
Z_{S_1,S_2} = 4 \sum_{nl\in S_1} \sum_{n'l'\in S_2} \sum_{m\in \{S_1,S_2\}} \frac{f_{nl} - f_{n'l'}}{\epsilon_{n'l'}-\epsilon_{nl}} \\ |\nabla_{nn'll'm}^z|^2 \delta(l\pm 1 - l') \Theta (\epsilon_{n'l'}-\epsilon_{nl})\,.
\end{multline}
In the above equations, $\nabla_{nn'll'm}^z$ is the $z$-component of the momentum integral matrix product,
\begin{equation}
    \nabla_{nn'll'm}^z = \langle \phi_{n'l'm} | \nabla_z | \phi_{nlm} \rangle\,,
\end{equation}
and $\Theta (\epsilon_{n'l'}-\epsilon_{nl})$ is the Heaviside step function. The derivation of the above expressions, and the expression for $\nabla_{nn'll'm}^z$ in terms of the radial KS orbitals and spherical harmonic functions, can be found in Appendix B.

\begin{figure}
    \centering
    \includegraphics[width=\columnwidth]{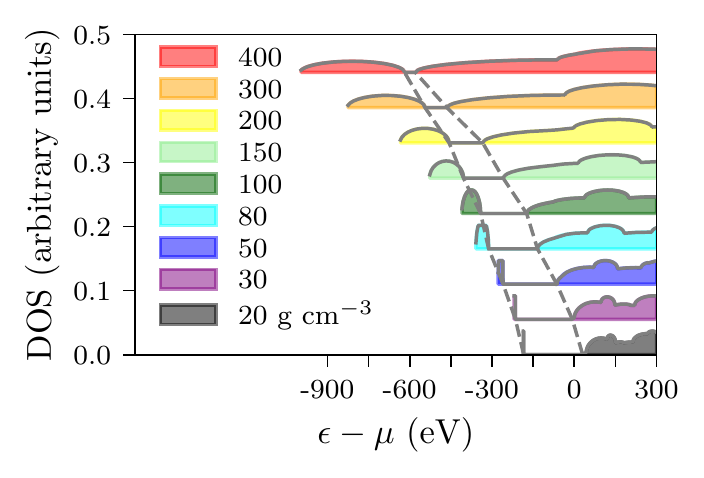}
    \caption{Density-of-states (DOS) with the band-structure AA model for Carbon at 100 eV and a range of densities. Orbitals with energies below the band-gap are assigned to the valence band, and orbitals with energies above the gap are assigned to the conduction band. Note that we deliberately mimic the style of Fig.~2 in Ref.\onlinecite{Redmer_Kubo_Greenwood_20}, so that readers can compare the DOS of the AA band-structure model to the DFT-MD results.}
    \label{fig:C_DOS}
\end{figure}

In a conventional AA model, unlike in plane-wave DFT calculations, there is no concept of a band-structure, which is problematic for determining
which subset of orbitals belongs to the conducting and valence bands.
We could, for example, use a threshold energy as the dividing line between conduction and valence electrons.
However, since we have the band-structure AA model at our disposal, we can use that to guide which orbitals belong in the conduction and valence bands. This is
just done manually (e.g. by inspecting the DOS, see Fig.~\ref{fig:C_DOS}).
Even when the conductivity is evaluated with the Dirichlet or Neumann boundary conditions, we use the band-structure model to determine the valence and conduction bands. 

In Fig.~\ref{fig:C_DOS}, we plot the DOS given by the AA band-structure model for Carbon at 100 eV and various densities. In this case, there is a clear valence band (to the left of the dotted lines) and conduction band (to the right). Through inspection of the energies, the valence band can be associated with the orbitals in the $1s$ band. Therefore, when evaluating the KG conductivity with the Dirichlet or Neumann condition, the $1s$ orbital is assigned to the valence band and all others to the conduction band. The same strategy is used in applications of the KG method in this paper. 

\begin{figure*}
    \centering
    \includegraphics{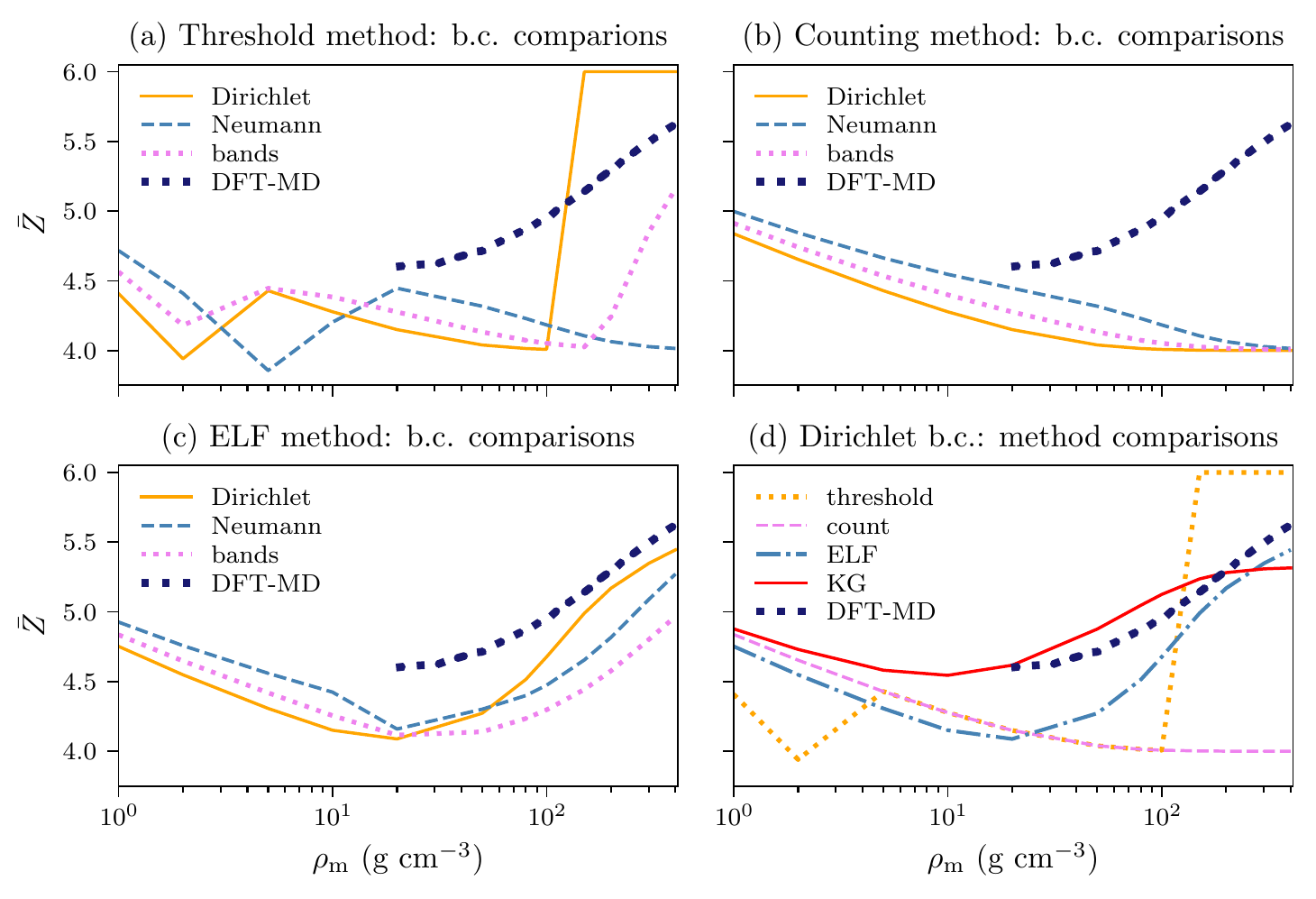}
    \caption{Comparison of different models for the mean ionization state (MIS) of Carbon with temperature 100 eV as a function of density. DFT-MD data is taken from Ref.~\cite{Redmer_Kubo_Greenwood_20} and uses the Kubo--Greenwood method. (a) Comparison of boundary conditions (b.c.s) with the MIS computed with the energy threshold method. (b) Comparison of boundary conditions with the MIS computed via the counting approach. (c) Comparison of boundary conditions with the MIS computed via the ELF. (d) Comparison of different methods for computing the MIS with the Dirichlet boundary condition.}
    \label{fig:C_DFT_comp}
\end{figure*}

\section{Results}

All calculations have been performed using the open-source average atom code atoMEC \cite{SciPy_atoMEC,atoMEC}. In Ref.~\onlinecite{SciPy_atoMEC}, we describe the structure of the code, together with general algorithmic and numerical details. Numerical details specific to this paper are given in the Supplemental Material \footnote{See Supplemental Material at [URL will be inserted by
publisher] for details regarding the computational implementation of the methods described in this paper, and all the code required to reproduce the results}. We note that the following libraries are used extensively by atoMEC: NumPy \cite{numpy}, SciPy \cite{scipy}, LIBXC \cite{libxc_2018}, mendeleev \cite{mendeleev2014}, and joblib \cite{joblib}.

In the following, we shall compare the four methods described for computing the MIS --- the canonical or ``threshold'' approach \eqref{eq:MIS_thresh}, the counting method \eqref{eq:Z_count}, the ELF method, and the KG approach \change{\eqref{eq:KG_Z_2}} --- against a higher fidelity DFT-MD benchmark and  experimental data. For the threshold, counting, and ELF results, we compare the Dirichlet and Neumann boundary conditions and the band-structure model \cite{Massacrier_bands_2021}. For the KG results, we use the Dirichlet boundary condition only. This is because the sum rule check for the total conductivity is observed very accurately (within $1\%$) across all conditions for the Dirichlet boundary condition, but not for the others. We use throughout the (spin-unpolarized) local density approximation (LDA) for the xc-functional \cite{PW_lda_1992}.
    
\begin{figure*}
    \centering
    \includegraphics{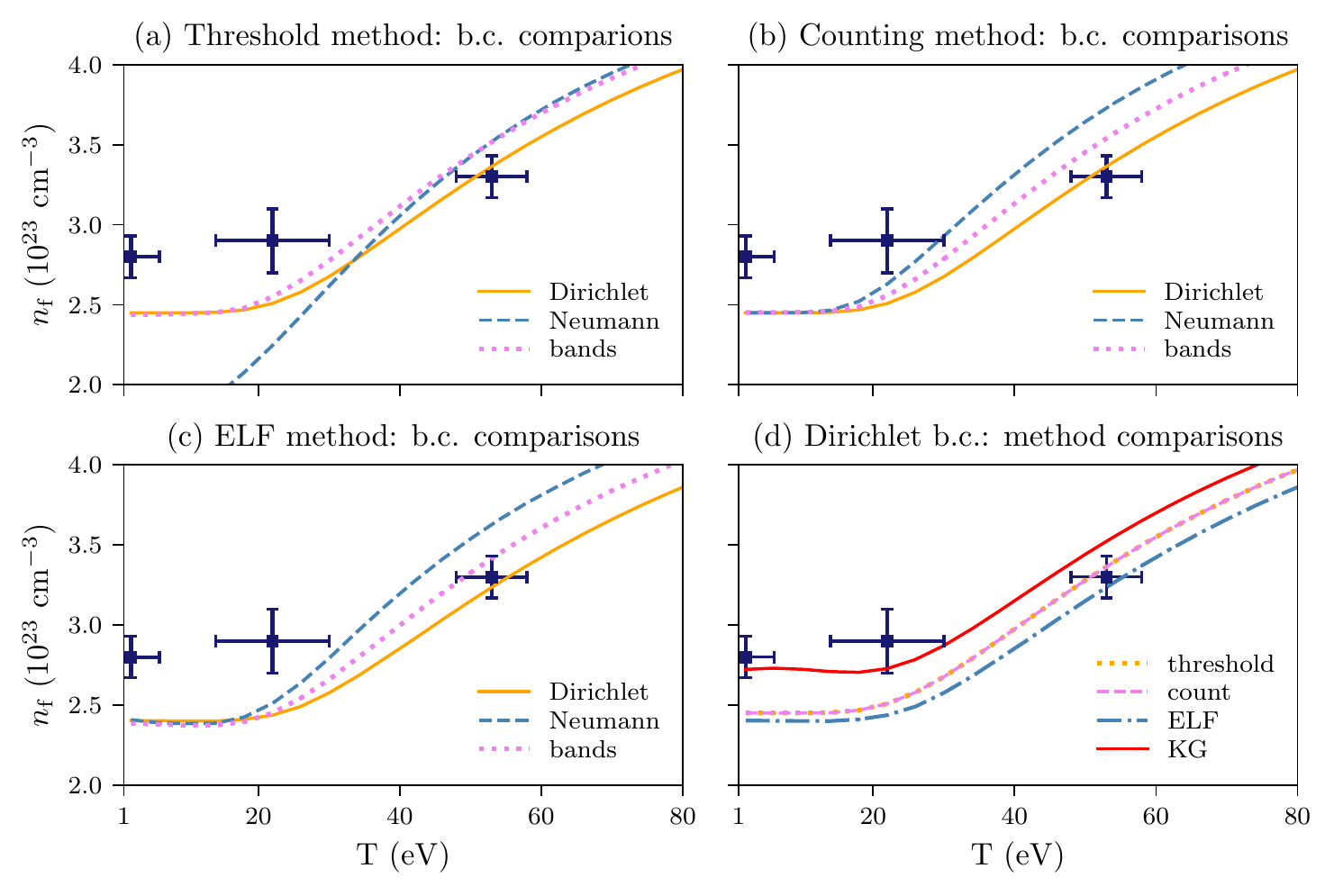}
    \caption{Comparison of different models and experimental data \cite{Glenzer_Beryllium_expt_2003} for the free electron density as a function of temperature for solid-density Beryllium ($\rho=1.85\ \textrm{g cm}^{-3}$). (a) Comparison of boundary conditions with the MIS computed with the energy threshold method. (b) Comparison of boundary conditions with the MIS computed via the counting approach. (c) Comparison of boundary conditions with the MIS computed via the ELF. (d) Comparison of different methods for computing the MIS with the Dirichlet boundary condition.}
    \label{fig:Be_expt_comp}
\end{figure*}

First, in Fig.~\ref{fig:C_DFT_comp}, we compare our results with the DFT-MD simulations for Carbon from Ref.~\onlinecite{Redmer_Kubo_Greenwood_20}. In Fig.~\ref{fig:C_DFT_comp}.~(a), we plot the MIS $\MIS$ using the canonical threshold method for each of the boundary conditions. We see that this method has severe limitations, especially at the highest densities, when the three boundary conditions yield completely different results. Furthermore, in the density-range in which the DFT-MD simulations were performed, none of the AA results are remotely close to the reference result.

In Fig.~\ref{fig:C_DFT_comp}.~(b), we plot the MIS using the counting method. In order to obtain these results, we took the electron density in the $1s$ orbital to be bound, and everything outside it to be free. In this case, we observe that the results are quite consistent between the different boundary conditions. However, they all tend to the wrong limit as the density increases. This is a consequence of the assumption that some orbitals --- in this case, the $1s$ orbital --- are bound states for the whole range of densities. Clearly, from both an intuitive and results-based perspective, this is not the case. Consider, for example, Fig.~\ref{fig:C_DOS}: as the material density increases, the part of the DOS that comes from the $1s$ orbital (to the left of the left dotted line) transforms in nature from a delta-like function (signifying bound electron density) to a wide energy band (signifying free electron density).

In Fig.~\ref{fig:C_DFT_comp}.~(c), we plot the MIS obtained via the ELF method against the DFT-MD benchmark. In order to obtain these results, we took the electron density in the $n=1$ sub-shell to be bound, and everything outside it to be free. We see that this approach yields a more realistic picture for the MIS, as the results from the three boundary conditions are at least consistent and capture the correct qualitative behaviour; however, they all systematically under-estimate the MIS relative to the DFT-MD result. Nevertheless, it is interesting that the ELF method, unlike the counting method, does go towards the correct limit as density increases. This suggests the ELF has some ability to inherently distinguish between components of electron density with different character.

Finally, in Fig.~\ref{fig:C_DFT_comp}.~(d), we compare results from the four methods (including KG) using the Dirichlet boundary condition with the DFT-MD simulation. Here we observe very strong agreement between our AA model and the DFT-MD benchmark for the KG result, until the highest densities at which point the KG result seems to have the wrong asymptotic behaviour. In this region, the ELF method actually appears to show better agreement with the reference result. Of course, the other limitation of the KG method is that it currently only works for the Dirichlet boundary condition, and it is possible that we would see inconsistencies between the boundary conditions, were a comparison possible.

In Ref.~\cite{Redmer_Kubo_Greenwood_20}, it was postulated that the AA result deviates from the DFT-MD result because the AA model does not account for the many-body interactions. Based on Fig.~\ref{fig:C_DFT_comp}, there is encouraging evidence that if the same theory is used to calculate the MIS for the AA and DFT-MD simulations, then the agreement is much better. Physically speaking, it is perhaps not unexpected that the KG result differs from the ELF and threshold approaches. After all, the KG conductivity is a frequency or time-dependent property, derived by considering the linear response of a system to a perturbation; on the other hand, the ELF and energy threshold are static properties. On that basis, we should not necessarily presume consistency between the different methods.

Next, we perform a similar set of comparisons for Beryllium in Fig.~\ref{fig:Be_expt_comp}, this time with fixed density equal to its ambient density ($\rho_\textrm{m}= 1.85\ \textrm{g cm}^{-3}$) from temperatures between $1-80\ \textrm{eV}$. This time, the benchmark results (shown as the 3 scattered points with error bars) are taken from an experiment, in which the free electron density $n_\textrm{f}$ was determined using X-ray scattering \cite{Glenzer_Beryllium_expt_2003}. The free electron density is directly related to the MIS,
\begin{equation}
    n_\textrm{f} = \frac{\MIS}{V},
\end{equation}
where $V = (4/3) \pi \RWS^3$ is the volume of the atom.
Like in the prior Carbon example, we have assumed under these conditions that the electron density in the $n=1$ (i.e. the $1s$ orbital) shell is bound, and everything outside it is free.

Again, the threshold results are shown in the top-left panel (a), the counting results in the top-right (b), and the ELF results in the bottom-left (c). This time, we see better agreement between the threshold results for the different boundary conditions, although the Neumann result is significantly different from the others at low temperatures. The counting and ELF results are somewhat similar, but resolve this inconsistency at low temperatures. Whilst all three techniques seem to capture roughly the right shape of the curve and agree quite well with the highest-temperature experimental measurement, they under-estimate the MIS for the lower-temperature results.

In Fig.~\ref{fig:Be_expt_comp}.~(d), we compare all three approaches for computing the MIS (threshold, ELF and KG) with just the Dirichlet boundary condition against the experimental data. Intriguingly, the KG results are in very close agreement with the lower temperature experimental results, although slightly over-predict the free electron density at the highest temperature. The KG result for the lowest temperature ($\tau\approx 2\ \textrm{eV}$) is particularly interesting, because it is the only method which correctly predicts the experimentally measured value of $\approx 2.8\times 10^{23}\ \textrm{cm}^{-3}$: this is higher than the value which we might naively expect if we take ambient density Beryllium to have two free electrons per atom, which corresponds to $n_\textrm{f}=2.45\times 10^{23}\ \textrm{cm}^{-3}$.
    


\begin{figure*}
    \centering
    \includegraphics{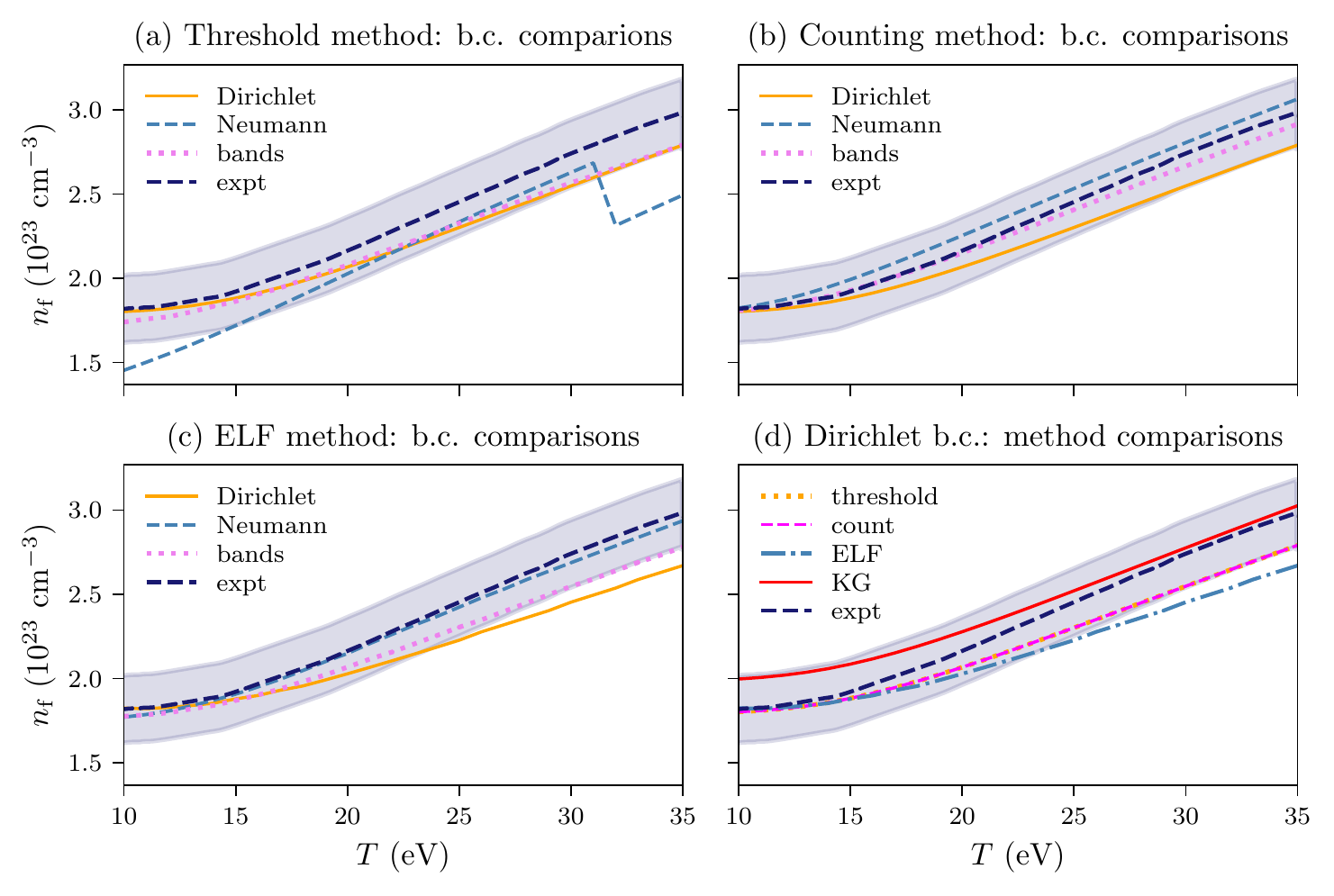}
    \caption{
    Comparison of different models and experimental data \cite{Vinko2015} for the free electron density as a function of temperature for ambient-density Aluminium ($\rho=2.7\ \textrm{g cm}^{-3}$). (a) Comparison of boundary conditions with the MIS computed with the energy threshold method. (b) Comparison of boundary conditions with the MIS computed via the counting approach. (c) Comparison of boundary conditions with the MIS computed via the ELF. (d) Comparison of different methods for computing the MIS with the Dirichlet boundary condition. The shaded region represents the experimental error bars.}
    \label{fig:Al_n_free}
\end{figure*}

\begin{figure*}
    \centering
    \includegraphics{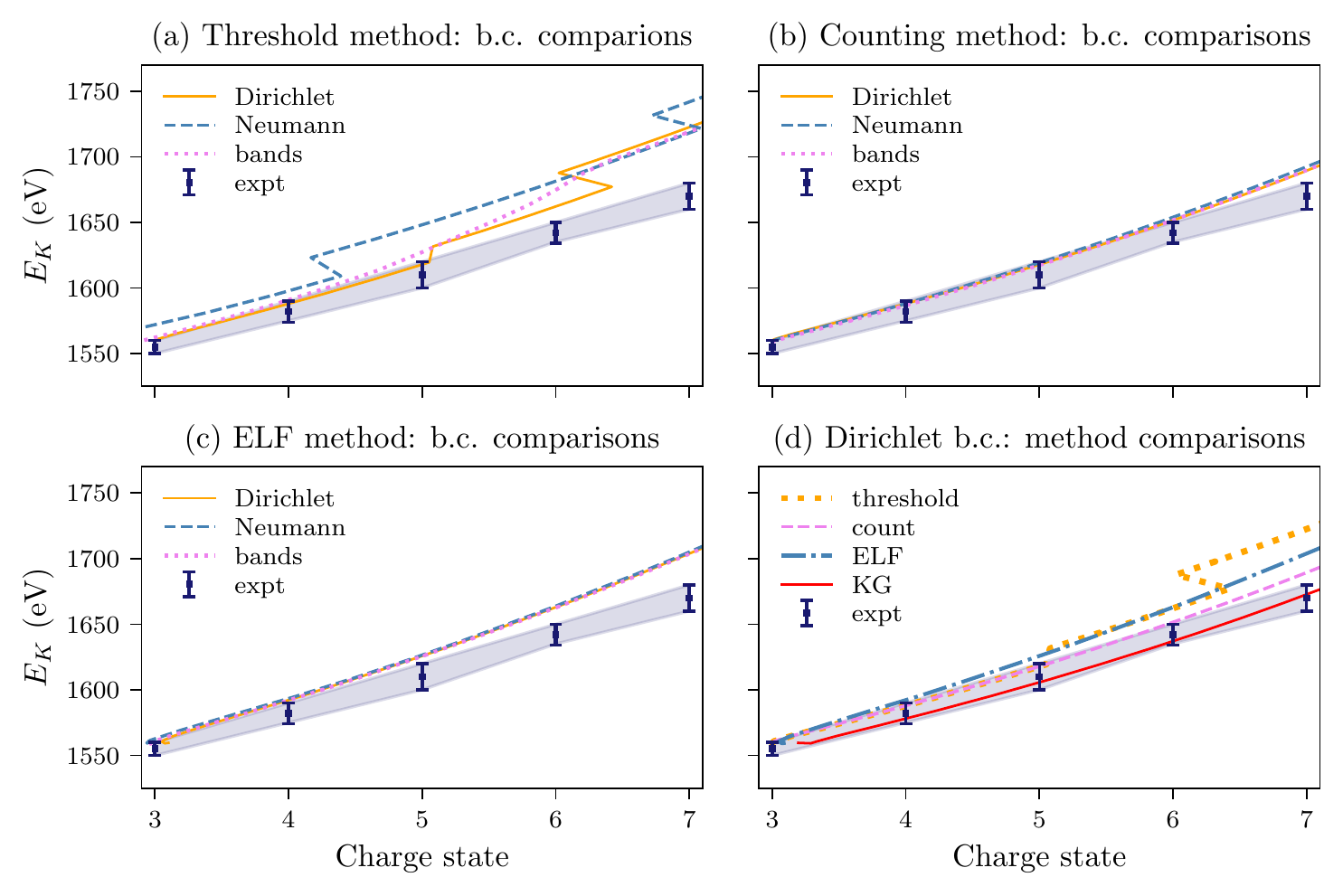}
    \caption{Comparison of different models and experimental data \cite{Aluminium_IPD_expt_2012} for the $K$-shell excitation energy as a function of the charge state (given by the MIS we compute) for ambient-density Aluminium ($\rho=2.7\ \textrm{g cm}^{-3}$). (a) Comparison of boundary conditions with the MIS computed with the energy threshold method. (b) Comparison of boundary conditions with the MIS computed via the counting approach. (c) Comparison of boundary conditions with the MIS computed via the ELF. (d) Comparison of different methods for computing the MIS with the Dirichlet boundary condition.}
    \label{fig:Al_K_exc}
\end{figure*}

The final comparisons we make are with a pair of experiments, both involving Aluminium at its ambient density ($2.7\ \textrm{g cm}^{-3}$). In the first experiment \cite{Vinko2015}, the free electron density $n_\textrm{f}$ and electron temperature were measured. We compare our AA results using the different methods and boundary conditions with the experimental data in Fig.~\ref{fig:Al_n_free}. In fact, under these conditions, the majority of the AA results actually lie within the experimental error bars, regardless of the method or boundary condition used. However, a notable exception is the prediction for $n_\textrm{f}$ given by the Neumann boundary condition with the threshold method (which has a sharp discontinuity at around 30 eV), further demonstrating the limitations of the threshold approach. Nevertheless, Fig.~\ref{fig:Al_n_free} indicates the AA model seems to be generally accurate under these conditions, independent of the method used to compute the MIS. \change{It appears that the ELF method with Neumann boundary condition is in particularly good agreement with the experimental benchmark in Fig.~\ref{fig:Al_n_free}.~(c). Since this is the only example to show such strong agreement, we prefer not to interpret this observation, but rather attribute it to chance.}


In the second experiment \cite{Aluminium_IPD_expt_2012}, the free electron density was not itself measured, but rather the $K$-shell ionization energy for different charge states. We use this data indirectly in the following way to compare our methods for calculating the MIS. For a range of temperatures between $1-100\ \textrm{eV}$, we compute the MIS and equate it to the charge state. We then take the $K$-edge ionization energy as the energy required to excite the $1s$ orbital to the continuum (with the continuum assumed to start at $\epsilon_\textrm{c}= v_\textrm{s}(\RWS)$ in our model). We also follow Ref.~\cite{Son_IPD_AA_2014} and shift the orbital energy by a constant equal to the difference between $\epsilon_{1s}-\epsilon_\textrm{c}$ and the experimentally measured $K$-shell ionization energy $E_K^0$ at zero temperature (1559.6 eV, \cite{thompson2001x}). Therefore the ionization energy is given by
\begin{gather}
    E_K = \epsilon_\textrm{c} - \epsilon_{1s}
+ \Delta E_K^0\,, \textrm{with}\\
 \Delta E_K^0 = (\epsilon_\textrm{c}^0 - \epsilon_{1s}^0) - E_K^0,
\end{gather}
where $\epsilon_\textrm{c}^0$ and $\epsilon_{1s}^0$ are the threshold energy and $1s$ orbital energy computed with the AA model at zero temperature. This shifting is necessary because it is well-known that KS-DFT systematically under-predicts ionization potentials using standard xc-functionals \cite{KronikIonization,NikNekionization}.

This comparison is shown in Fig.~\ref{fig:Al_K_exc}. In Fig.~\ref{fig:Al_K_exc}.~(a), we again see several discontinuities in the threshold results (for the Neumann and Dirichlet conditions), and a systematic deviation from the experimental results for higher charge states. The reason for these discontinuities, as discussed in detail in Ref.~\cite{callow_first_principles}, is because the KS orbital energies are temperature-dependent; if an orbital crosses the energy threshold at a certain temperature then it will change its classification from bound to free (or vice-versa) and the MIS will change instantaneously. An advantage of the band-structure model is that it is not prone to discontinuities in the MIS as a function of temperature, as can be seen in Fig.~\ref{fig:Al_K_exc}. This is because occupations of the non-core states in the band-structure model are spread across a band: as the limits of the energy band change, the MIS smoothly changes. This demonstrates a significant advantage of the band-structure model when the threshold method is used.

In Fig.~\ref{fig:Al_K_exc}.~(b), the counting approach is seen to yield consistent agreement, both internally between the different boundary conditions, and with the experimental benchmarks. For most charge states, the results lie just inside the experimental error, with some deviation seen as the charge state goes above 6 (corresponding to temperatures $\gtrapprox 60$ eV). In Fig.~\ref{fig:Al_K_exc}.~(c), the ELF results are self-consistent between the boundary conditions but also display the same systematic deviation from the experimental data. This is likely a result of the MIS being under-estimated by these methods, as we have seen for the previous examples. However, in Fig.~\ref{fig:Al_K_exc}.~(d), we observe that the KG results lie consistently within the experimental range.  This shows further promise that the KG approach for calculating the MIS agrees very well with experimental measurements.

\section{Summary and discussions}

In this paper, we have explored different ways of computing the mean ionization state (MIS) --- an essential property in warm dense matter and dense plasmas --- using a KS-DFT average-atom model. Following comparisons of the different methods with DFT-MD results and experimental data, we summarize the main findings of our paper below.

The canonical method for computing the MIS, which partitions the orbitals into bound and free states based on their energies, is generally insufficient. It often causes unphysical discontinuities, and inconsistencies between different boundary conditions. If it is to be used, it is much safer to do so with the band-structure model \cite{Massacrier_bands_2021}, since this avoids (at least as a function of temperature) the discontinuities.

We have explored an approach which we call the `counting' method (\change{which was also used for the non-average-atom XCRYSTAL model in Ref.~\cite{xcrystal}}), where the orbitals are partitioned into bound and free states based on some pre-defined intuition. This does not suffer from the discontinuities present in the threshold method, and also yields consistent results between the boundary conditions. However, it breaks down when orbitals cannot be \emph{a priori} identified as being strictly bound or free in character.

We have developed an approach which uses the electron localization function (ELF) to partition the orbitals. Like the counting method, this requires a choice by the user as to which shells should be considered bound or free; however, the shells in this case do not necessarily correspond directly to particular orbitals, and so it yields better results than the counting method when the material density is varied.

We have applied a method which uses the Kubo--Greenwood conductivity \cite{Redmer_Kubo_Greenwood_20} to our average-atom model. This also requires a choice by the user regarding a separation of orbitals into valence and conducting bands, but the resulting MIS has a sophisticated non-linear dependence on this separation. This seems to yield the strongest agreement with DFT-MD and experimental benchmarks. However, so far we have applied it only to the Dirichlet boundary condition, since sum rules are not satisfied for the other boundary conditions.

Roughly speaking, we observe two different physical situations in this paper. In one instance, Figs.~\ref{fig:Be_expt_comp}, \ref{fig:Al_n_free} and \ref{fig:Al_K_exc}, the temperature is varied for a metallic material whose mass density is fixed to its ambient value. This case is relatively straightforward: with the exception of the canonical approach with the Dirichlet and Neumann boundary conditions, all the methods yield good agreement with the benchmark data. This is because, for metals under a wide range of temperatures, the core orbitals do not undergo much change in character so can always be treated as bound states.

The other instance, Fig.~\ref{fig:C_DFT_comp}, in which the material density (in this paper, Carbon) is varied at fixed temperature, is far more challenging. Neither the threshold or counting method is sufficient in this case; however, both the ELF and KG methods yield promising results.

It is worth noting that the KG approach has a fundamental difference compared to the other methods, since it is based on a dynamic rather than static theory. Empirically, it seems to yield systematically higher predictions for the MIS than the other methods, and also seems closer to the experimental benchmarks. This perhaps follows from the technique used to determine the free electron density in such experiments. 

Based on the previous point, it may be that the ``best'' method to compute the MIS depends on what is desired. If the aim is to compare or provide data for an experimental fitting, the KG approach would appear to be the best approach. However, it may be that for other purposes, such as when the MIS is used as input for hydrodynamics codes, alternative methods could be favourable. This point will benefit from further investigation in future.

As a final comment, we note that more experimental data would help identify which method is most accurate across the widest range of conditions. However, high-quality experimental measurements of the free electron density (or MIS) are not trivial to come by. The assumptions used to calculate the MIS --- for example, from the ratio of the inelastic to elastic scattering in X-ray scattering experiments \cite{Glenzer_Beryllium_expt_2003} --- may be more likely to break down under the ``harder'' case of a material whose density is varied, as described earlier. This presents a major challenge for bench-marking different approaches for calculating the MIS.

In summary, the methods and data we have presented in this paper should indicate when certain methods for computing the MIS in average-atom models work, and when they might be expected to break down. With two of the methods --- the ELF and KG approaches --- the results are promising for all the examples we have tested. This is of particular interest because our AA code can typically run on a laptop in the time-scale of minutes --- far less computationally demanding than DFT-MD simulations.



\section*{Acknowledgements}

We thank G\'erard Massacrier for constructive discussions regarding the band-structure AA model; Martin French for a useful discussion about experimental measurements of the MIS; Kieron Burke and Maximilian Schörner for insightful comments regarding definitions of the MIS; and particularly the anonymous referee for suggesting the counting method. We are also grateful to the organizers of the ``Average atom models for warm dense matter workshop'' at UC Berkeley in June 2021, which motivated the idea for this paper. This work was partially supported by the Center for Advanced Systems Understanding (CASUS) which is financed by Germany’s Federal Ministry of Education and Research (BMBF) and by the Saxon state government out of the State budget approved by the Saxon State Parliament. EK greatly appreciates the support of the Alexander von Humboldt Foundation.

\begin{appendices}

\section{Derivation of $w_k$ terms in band-structure model}

The energy integral to compute the density in the band-structure model, Eq.~\eqref{eq:dens_masac}, must be discretized in practise. It therefore
becomes a summation over energies within each band which we now denote
by index \(k\),

\begin{align}
\nonumber
 n(r) &= 2\sum_{knl}(2l+1) \delta\epsilon_{knl} g_{knl}(\epsilon_{knl},\epsilon_{nl}^\pm) \\ &\hspace{6em} \times f_{knl}(\epsilon_{knl},\mu,T) |X_{knl}(r)|^2\,,\\
 g_{knl}&(\epsilon_{knl},\epsilon_{nl}^\pm) =\frac{8}{ \pi \Delta_{nl}^2} \sqrt{(\epsilon^+_{nl}-\epsilon_{knl})(\epsilon_{knl} - \epsilon^-_{nl})}\,.
\end{align}

We now simplify the above expressions, because
this simplification was not discussed in the original paper. Firstly, we
note that the energy spacing in the discretization of the energy band
\(\delta\epsilon_{knl}\) is therefore given by
\begin{equation}
\delta\epsilon_{knl} = \frac{\epsilon^+_{nl} - \epsilon^-_{nl}}{N_k-1} = \frac{\Delta_{nl}}{N_k-1}\,,
\end{equation}
where \(N_k\) is the number of \(k\) points (the denominator is equal to
\(N_k-1\) because there are \(N_k-1\) spacings for \(N_k\) total
points). The product
\(\delta\epsilon_{knl} \times g_{knl}(\epsilon_{knl},\epsilon_{nl}^\pm)\)
therefore can be written as
\begin{multline}
\delta\epsilon_{knl} \times g_{knl}(\epsilon_{knl},\epsilon_{nl}^\pm) =\\ \frac{8}{\pi\Delta_{nl}(N_k-1)} \sqrt{(\epsilon^+_{nl}-\epsilon_{knl})(\epsilon_{knl} - \epsilon^-_{nl})}\,.
\end{multline}

Next, we note that the energies in a band \(\epsilon_{knl}\) can be
re-written as
\begin{align}
\epsilon_{knl} &= \epsilon_{nl}^- + \frac{k}{N_k-1}\Delta_{nl} \\
               &= \epsilon_{nl}^+ + \frac{k-(N_k-1)}{N_k-1}\Delta_{nl}
\end{align}
Substituting the above expressions into the product
\(\delta\epsilon_{knl} \times g_{knl}(\epsilon_{knl},\epsilon_{nl}^\pm)\)
leads to the following expression:
\begin{equation}
\delta\epsilon_{knl} \times g_{knl}(\epsilon_{knl},\epsilon_{nl}^\pm) = \frac{8}{\pi(N_k-1)^2}\sqrt{k(N_k-1-k)}\,.
\end{equation}
It is clear the above equation is in fact independent of the quantum
numbers \(n\) and \(l\). The density \(n(r)\) thus becomes
\begin{gather}
n(r) = 2\sum_{k} w_k \sum_{nl}(2l+1) f_{knl}(\epsilon_{knl},\mu,T) |X_{knl}(r)|^2\,,\\
w_k = \frac{8}{\pi(N_k-1)^2}\sqrt{k(N_k-1-k)}\,.
\end{gather}

\section{Kubo--Greenwood conductivity in the average-atom model}

In the spherically symmetric case, the KS orbitals are expanded in the form $\phi_i(\vec{r})=\phi_{nlm}(r,\theta,\phi)=\rnl Y_l^m(\theta,\phi)$, and the KG conductivity \eqref{eq:kg} becomes

\begin{multline}
\sigma_{S_1,S_2}(\omega) = \frac{2\pi}{3V\omega} \sum_{nlm\in S_1} \sum_{n'l'm'\in S_2} (f_{nlm} - f_{n'l'm'})\\ |\langle \phi_{nlm} | \nabla | \phi_{n'l'm'} \rangle|^2 \delta (\epsilon_{n'l'm'} - \epsilon_{nlm} - \omega)\,.
\end{multline}
Note that, in the band-structure model, this becomes
\begin{multline}
\sigma_{S_1,S_2} (\omega) = \frac{2\pi}{3V\omega} \sum_k w_k \sum_{nlm\in S_1} \sum_{n'l'm'\in S_2} (f_{knlm} - f_{kn'l'm'}) \\|\langle \phi_{knlm} | \nabla | \phi_{kn'l'm'} \rangle|^2 \delta (\epsilon_{kn'l'm'} - \epsilon_{knlm} - \omega)\,,
\end{multline}
similar to the KG conductivity in plane-wave DFT codes. For simplicity,
and because we only use the KG conductivity with
Dirichlet boundary condition in this paper, we shall present the equations without the
\(k\)-index. Since the summation only involves orbitals with the same
\(k\)-value, it is straightforward to re-introduce this at the end of
the derivation.

We focus first on the integral component of the equation for
\(\sigma(\omega)\), which is given by
\begin{align}
&|\langle \phi_{nlm} | \nabla | \phi_{n'l'm'} \rangle|^2 \nonumber
\\&\hspace{3em}= \sum_{i=1}^3 \langle \phi_{nlm} | \nabla_i | \phi_{n'l'm'} \rangle \langle \phi_{n'l'm'} | \nabla_i | \phi_{nlm} \rangle \label{eq:kg_sph_1} \\
&\hspace{3em}=3 \langle \phi_{nlm} | \nabla_z | \phi_{n'l'm'} \rangle \langle \phi_{n'l'm'} | \nabla_z | \phi_{nlm} \rangle\,, \label{eq:kg_sph_2}
\end{align}
where the second equation \eqref{eq:kg_sph_2} follows from \eqref{eq:kg_sph_1} because the contribution from each
cartesian component of the gradient is identical in spherically
symmetric systems. We choose the \(z\) component because, in the
traditional transformation between cartesian and spherical co-ordinates,
this leads to a simpler set of equations. Let us now focus on the
following term,
\change{\begin{align} \label{eq:mel_kg}
\langle \phi_{n'l'm'} | \nabla_z | \phi_{nlm} \rangle \
&= \nabla_{nn'll'mm'}^z \\
&=R^{(d)}_{nn'll'} P^{(2)}_{lml'm'} \delta_{mm'} \nonumber \\
&\hspace{3em}+ R_{nn'll'} P^{(4)}_{lml'm'} \delta_{mm'},
\end{align}}
which has been taken from Ref. \onlinecite{Trickey_Kubo_Greenwood_2017}.
We do not derive the above expression, but instead direct readers to the
aforementioned paper where it is derived in full.

The components of the matrix element \eqref{eq:mel_kg} are given by
\change{\begin{align}
R^{(d)}_{nn'll'} &= 4\pi\int_0^{R_\textrm{VS}} \textrm{d}r r^2 X_{n'l'}(r) \frac{\textrm{d}X_{nl}(r)}{\textrm{d}r}\\
 R_{nn'll'} &= 4\pi\int_0^{R_\textrm{VS}} \textrm{d}r r X_{n'l'}(r) X_{nl}(r) \\
 P^{(2)}_{lml'm'} &= 2\pi C_{lm}C_{l'm'} \int_{-1}^{1} \textrm{d}x x P_{l'}^{m'} (x) P_{l}^{m}(x) \\
 P^{(4)}_{lml'm'} &= -2\pi C_{lm}C_{l'm'} \nonumber \\ &\hspace{4em}\int_{-1}^{1} \textrm{d}x (1-x^2) P_{l'}^{m'} (x) \frac{\textrm{d}P_l^m(x)}{\textrm{d}x}\\
 C_{lm} &= \sqrt{\frac{2l+1}{4\pi}}\sqrt{\frac{(l-|m|)!}{(l+|m|)!}}\,,
\end{align}}
where \(P_l^m(x)\) are the Legendre polynomials. Note there are some
additional factors of \(4\pi\) in the above expressions compared to Ref.
\onlinecite{Trickey_Kubo_Greenwood_2017}, due to different conventions in
normalization of the orbitals.

Returning to the expression for
\(\sigma(\omega)\), we now have
\begin{multline}
\sigma_{S_1,S_2}(\omega) = \frac{2\pi}{V\omega} \sum_{nl\in S_1} \sum_{n'l'\in S_2} \sum_{m\in \{S_1,S_2\}} (f_{nl} - f_{n'l'}) \\ |\nabla_{nn'll'm}^z|^2 \delta (\epsilon_{n'l'} - \epsilon_{nl} - \omega) \delta(l\pm 1 - l')\,.
\end{multline}
In the above, the double summation over \(m\) has been reduced to a
single summation because of the presence of the the \(\delta_{mm'}\) in
\(\nabla_{nn'll'mm'}^z\). Additionally, the \(\delta(l\pm 1 - l')\)
comes from sum rules in the evaluation of the \(P^{(2,4)}\) integrals.

Given the relationship between the conductivity and the number of
electrons \(Z_{S_1,S_2}\) (Eqs. \ref{eq:KG_Z_1} and \ref{eq:KG_Z_2}), we recover the following expression for $Z_{S_1,S_2}$,
\begin{multline}
Z_{S_1,S_2} = 4 \sum_{nl\in S_1} \sum_{n'l'\in S_2} \sum_{m\in \{S_1,S_2\}} \\ \frac{f_{nl} - f_{n'l'}}{\epsilon_{n'l'}-\epsilon_{nl}}|\nabla_{nn'll'm}^z|^2 \delta(l\pm 1 - l') \Theta (\epsilon_{n'l'}-\epsilon_{nl})\,.
\end{multline}

\end{appendices}

\bibliography{main}

\end{document}